\documentclass[final,times,twocolumn]{elsarticle}
\usepackage{amssymb,amsmath,hyperref,url,marginnote,xcolor}
\biboptions{sort&compress}
\journal{Physics Letters B}
\allowdisplaybreaks[2]

\usepackage{graphicx}
\usepackage{subcaption}

\newcommand{\df}{\mathrm{d}}

\newcommand{\pt}{p_{\mathrm{T}}}
\newcommand{\ptg}{p_{\mathrm{T}}^{\gamma}}
\newcommand{\ptjet}{p_{\mathrm{T}}^{\mathrm{jet}}}
\newcommand{\xjg}{x_{j\gamma}}
\newcommand{\PbPb}{\text{PbPb}}
\newcommand{\pp}{\text{pp}}
\newcommand{\JEWEL}{\textsc{Jewel}}
\newcommand{\pTsub}{p_{\mathrm{T}}^{\mathrm{sub}}}

\usepackage[normalem]{ulem}

\begin{document}

\begin{frontmatter}

\affiliation[nikhef]{organization={Nikhef},
              addressline={Science Park 105}, 
              postcode={1098 XG}, 
              city={Amsterdam}, 
              country={The Netherlands}}

\affiliation[uu]{organization={Utrecht University},
              addressline={P.O. Box 80000},
              postcode={3508 TA},
              city={Utrecht},
              country={The Netherlands}}

\affiliation[uva]{organization={Institute of Physics, University of Amsterdam},
              addressline={Science Park 904}, 
              postcode={1098 XH}, 
              city={Amsterdam},
              country={The Netherlands}}

\title{Using $\gamma$+jets to quantify medium-induced jet broadening in heavy-ion collisions}

\author[nikhef]{Ankita Budhraja}
\author[nikhef,uu]{Marco van Leeuwen}
\author[nikhef,uva]{Wouter J.~Waalewijn}

\begin{abstract}
The structure of jets produced in high-energy nucleus-nucleus collisions carries information on parton energy loss and interaction in the quark-gluon plasma. This parton energy loss results in migration of jets in terms of transverse momentum, leading to a selection bias in inclusive jet measurements. 
Using the \JEWEL\, event generator, we investigate a strategy to reduce selection bias and access medium-induced jet broadening, in an experimentally viable way. 
As a baseline, we first consider large-$R$ $\gamma$+jet events, with little selection bias, which show a clear signal of medium-induced broadening in the jet girth.
However, large-$R$ is experimentally inaccessible due to the large underlying event fluctuations in heavy-ion collisions, so we re-cluster a collection of small-radius ($r_{\rm sub} = 0.2$) jets as a proxy for the large-radius ($R = 1.2$) jet, which we refer to as a trimmed jet.
We quantify to what extent trimmed jets recover signals of jet broadening and investigate its dependence on the subjet cone size and the minimum transverse momentum.
We study the internal structure of the subjets to expose the dependence of jet quenching on different subjet configurations, finding a strong narrowing of $\PbPb$ jets for the 1-subjet configurations, and a visible signature of medium-induced broadening for sub-leading subjets. 
The jet radial profile reveals that contributions from medium-induced broadening are distinct from radiation in pp collisions. These contributions are not easily recovered using our trimming procedure, but are substantially enhanced in single-subjet configurations when considering the profile \emph{beyond} the subjet radius.
This provides guidance for future experimental studies of jet-medium interactions using $\gamma$+jet in heavy-ion collisions.
\end{abstract}


\end{frontmatter}

\section{Introduction}
Jets produced in ultra-relativistic nuclear collisions at the Large Hadron Collider (LHC) and Relativistic Heavy-Ion Collider (RHIC) serve as calibrated probes of the quark-gluon plasma (QGP), a deconfined state of strongly-interacting matter formed in the aftermath of these collisions~\cite{Busza:2018rrf,Cao:2020wlm,Apolinario:2022vzg,Cunqueiro:2021wls,Arslandok:2023utm,Wang:2025lct}. Hard partons produced in the initial scattering traverse the QGP and interact with it through radiative and elastic processes, losing energy and acquiring transverse momentum kicks from scattering centers in the medium~\cite{Baier:1996sk,Zakharov:1996fv,Gyulassy:2000fs,Wang:2001ifa,Wiedemann:2000za}. These interactions modify the angular and energy distribution of the ensuing parton shower, imprinting characteristic signatures on the final-state particles. 

Jet substructure observables that encode the angular distribution of energy inside a jet offer a direct window into the microscopic properties of the QGP~\cite{Mehtar-Tani:2016aco,Caucal:2019uvr,Casalderrey-Solana:2014bpa,Milhano:2017nzm,Apolinario:2022vzg,Pablos:2022mrx,Cunqueiro:2021wls,Andrews:2018jcm}. However, determining these properties from inclusive jet substructure measurements has so far remained complicated by the \emph{jet fragmentation bias}, i.e., jets selected at fixed reconstructed $\ptjet$ in collisions of lead nuclei ($\PbPb$) at LHC are enriched in those that lost comparatively little energy to the QGP. The jets that are more quenched by medium interactions migrate to lower reconstructed $\ptjet$ and fall outside the selection window, so any comparison to the $\pp$ baseline at the same $\ptjet$ conflates genuine QGP modification with a selection-driven bias toward less quenched jets~\cite{Brewer:2021hmh,Cunqueiro:2023vxl}. This bias produces an \emph{apparent narrowing} of the jet radiation pattern observed experimentally across a range of substructure observables~\cite{CMS:2013lhm,ALICE:2017nij,CMS:2018zze,ALICE:2018dxf,ALICE:2019qyj,ALargeIonColliderExperiment:2021mqf,ALICE:2023dwg,ALICE:2024jtb}, even though the medium interactions are expected to broaden the perturbative jet shower~\cite{Baier:2000mf,Gyulassy:2003mc,Kovner:2003zj,Ovanesyan:2011xy,Vaidya:2020cyi}. 
In this work, we propose and study an experimentally viable strategy to expose the medium-induced jet broadening by using photon-jet events to reduce the selection bias effects.

The photon-tagged channel partially circumvents the jet fragmentation bias by using the photon transverse momentum $\ptg$ as the baseline to determine the $\pt$ of the recoiling jet~\cite{Wang:1996yh,CMS:2018mqn,ATLAS:2023iad}. As the photon carries no color charge, it remains unaffected by the QGP and provides an anchor for the initial parton momentum. Moreover, the momentum imbalance 
\begin{equation}
    \xjg = \frac{\ptjet}{\ptg}\, ,
\end{equation}
directly measures the fractional momentum retained by the jet after medium interaction. However, a residual bias still survives that now depends on the selection made on $\xjg$: a tight selection preferentially retains less quenched events, while a more inclusive selection reduces this effect retaining more quenched jets~\cite{Wang:2024plm}. The impact of this bias on substructure measurements was demonstrated compellingly by CMS~\cite{CMS:2024zjn}, which found that the sign of jet structure modification in $\gamma$+jet events depends on the $\xjg$ threshold: a tight selection ($\xjg > 0.8$) yields jet narrowing relative to $\pp$, while a more inclusive selection ($\xjg > 0.4$) reduces this effect. Certain theoretical studies~\cite{Luo:2018pto,Tachibana:2025rcx,Wang:2024plm} have also confirmed $\gamma$+jets to provide a significantly less biased probe of medium-induced angular modifications than inclusive jets.

\begin{figure}[!htb]
    \centering
    \includegraphics[scale=0.5]{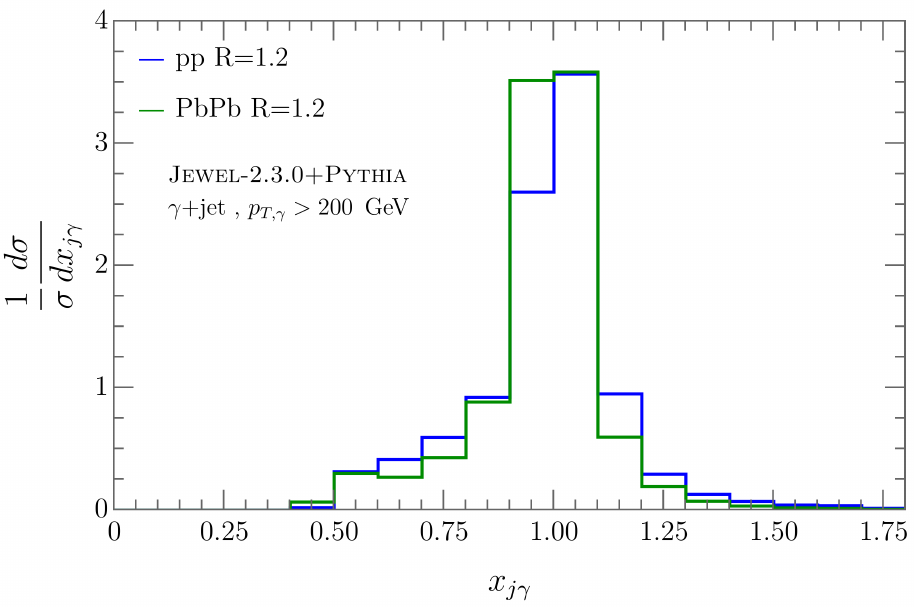}
    \caption{The momentum imbalance, $\xjg$, between the recoiling jet and the photon, computed over the constituents of the leading large $R=1.2$ jet for $\pp$ and $\PbPb$. The $\xjg$ distributions are similar in $\pp$ and $\PbPb$.}
    \label{fig:xjg_largeR}
\end{figure}

\begin{figure}[!htb]
    \centering
    \includegraphics[scale=0.48]{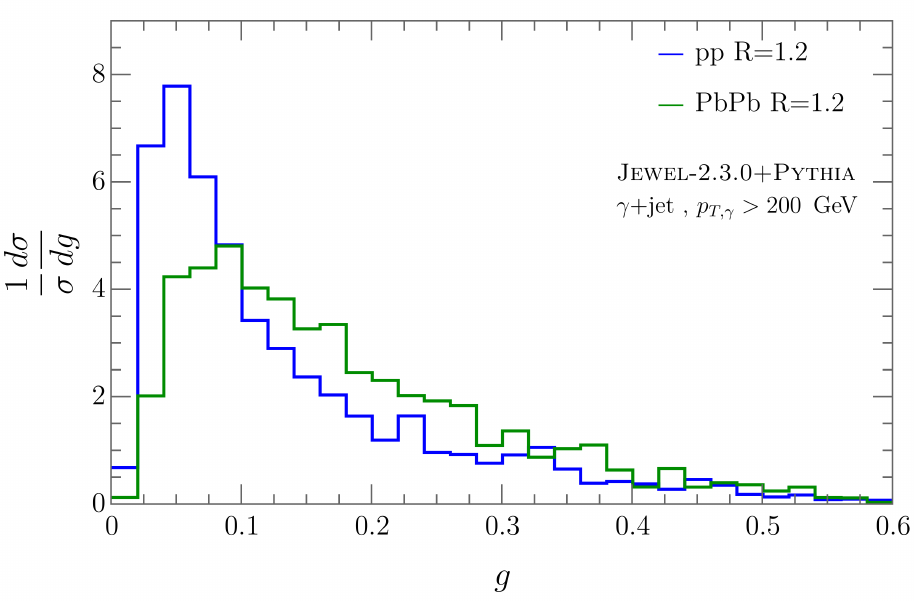}
    \caption{ The jet girth $g$ distribution of jets in $\pp$ and $\PbPb$ for $R=1.2$ jets recoiling against the photon with $\ptg > 200$ GeV. The large tail of the distribution for large $g$ values presents a strong signal of medium-induced broadening in $\PbPb$ jets.}
    \label{fig:girth_largeR}
\end{figure}

A complementary approach to suppress the fragmentation bias is to enlarge the jet reconstruction radius $R$. Large-$R$ jets recover a greater fraction of energy transported to large angles through jet-medium interactions
~\cite{Pablos:2019ngg,ALICE:2023waz,2013220}, restoring the energy balance between the jet and the photon and rendering the $\xjg$ distribution nearly insensitive to the overall energy loss, as shown in Fig.~\ref{fig:xjg_largeR} for $R=1.2$. From this Figure, we observe that the recoiling $R=1.2$ jet is nearly balanced with the photon in both $\pp$ and $\PbPb$, and the $\xjg$ distributions are almost identical to each other. This minimizes out-of-cone radiation, maximizing the opportunity to study medium effects from the in-cone radiation.
For instance, when comparing the transverse spread (\textit{jet girth} $g$) of the large $R=1.2$ jet in $\pp$ and $\PbPb$, as shown in Fig.~\ref{fig:girth_largeR}, in \JEWEL{} simulations, we observe a clear indication that in-medium jets tend to be broader than jets in vacuum.\footnote{We have verified that this broadening of the girth distribution in $\PbPb$ is stable to the background subtraction procedure in \JEWEL~\cite{KunnawalkamElayavalli:2017hxo,Milhano:2022kzx}.} However, the downside of this approach is that it is not experimentally realizable, since large-$R$ jets are challenging to measure in heavy-ion collisions due to the presence of fluctuations in the large underlying event~\cite{ATLAS:2023hso}. 

In this Letter, we study an experimentally viable strategy to probe jet-medium interactions with reduced selection bias effects. Using a procedure introduced in Ref.~\cite{Nachman:2014kla} and utilized by ATLAS for inclusive jets in heavy-ion collisions~\cite{ATLAS:2023hso}, we generate a proxy of the large $R=1.2$ jet by re-clustering a collection of small-radius, $r_{\rm sub}$, subjets. 
Since the minimum transverse momentum cut $\pTsub$ applied to the small-$r_{\rm sub}$ subjets serves as a natural grooming criterion, we refer to this as \textit{jet trimming}~\cite{Krohn:2009th}. The trimmed large $R=1.2$ jet is much less sensitive to background contamination~\cite{Zhang:2021sua} and can be experimentally realized.

Using the \JEWEL{} Monte Carlo generator, we first quantify the degree to which this trimming recovers the physics of direct large-$R$ jets in $\gamma$+jet events, both in terms of the momentum imbalance $\xjg$ and jet substructure. We focus on the \textit{jet girth} $g$ distribution~\cite{CATANI1992269,Dokshitzer:1998kz} of the trimmed jet and investigate its dependence on the cone size, $r_{\rm sub}$, and  minimum transverse momentum cut $\pTsub$. 

The `jet trimming' technique also exposes a number of new jet substructure observables for configurations with multiple hard subjets. The corresponding color charges may be independently resolved by the medium, leading to stronger energy loss than configurations containing only a single resolved subjet~\cite{Casalderrey-Solana:2012evi}. 
This suppression of energy loss as a function of the number of resolved prongs in a jet was observed experimentally by ATLAS~\cite{ATLAS:2023hso}, reported in terms of the nuclear modification factor $R_{\rm AA}$ for these configurations. 

We explore how the jet girth distribution depends on the structure of the subjet configurations forming the large-$R$ jet. Our findings reveal that while the leading jet in the configurations with a single-subjet structure shows narrowing, the sub-leading subjet in the 2-subjet structures exhibits clear signals of jet broadening.  Using the jet radial profile~\cite{Seymour:1997kj}, we find that the radiation responsible for broadening in Fig.~\ref{fig:xjg_largeR} is at larger angles than the radius of the subjet.

The rest of this Letter is organized as follows: in Sec.~\ref{sec:simulation} we describe the simulation details and jet observables explored in this study. In Sec.~\ref{sec:reclustering} we elaborate our jet trimming procedure and show to what extent the trimmed small-$r_{\rm sub}$ jets recover the physics contained in the large-$R$ jets. In Sec.~\ref{sec:subjets}, we present a differential analysis of the subjets forming the large-$R$ jet. We summarize in Sec.~\ref{sec:summary}. In~\ref{sec:selection}, we study the impact of varying the $\xjg$ cut on the jet substructure distribution.

\section{Analysis set-up and Jet Observables}
\label{sec:simulation}

\subsection{Simulation Details}
The samples for pp and PbPb jet events used in this work were generated with {\textsc{Jewel}}-2.3~\cite{Zapp:2012ak,Zapp:2013vla,KunnawalkamElayavalli:2016ttl}. {\textsc{Jewel}} is a Monte Carlo event generator that simulates jet evolution both in vacuum and in a thermal medium, modeling the QCD evolution of hard partons and their scatterings with the medium in a perturbative framework based on \textsc{Pythia} 6~\cite{Sjostrand:2006za}. The final-state parton shower is interleaved with medium scatterings such that vacuum-like and medium-induced emissions are treated within the same perturbative framework. In {\textsc{Jewel}}, the relative contributions of elastic and inelastic processes in the medium are described using a virtuality-ordered shower at leading logarithmic accuracy~\cite{Zapp:2012ak}. 

{\textsc{Jewel}} implements destructive interference between emissions from successive medium scatterings through a probabilistic formulation of the QCD analog of the Landau--Pomeranchuk--Migdal (LPM) effect~\cite{Zapp:2011ya}. In this formulation, all emissions are initially allowed, but the one with the longest formation time is discarded. The expansion of the medium is modeled through a boost-invariant longitudinal expansion of an ideal quark--gluon gas~\cite{PhysRevD.27.140}. We work with this default medium model throughout.

Among other medium effects, part of the energy deposited by the jet in the medium via momentum exchanges with the medium partons, can end up being reconstructed as final-state jet constituents. This effect is usually referred to as \textit{medium response}~\cite{Cao:2022odi}. By default, \JEWEL{} does not store the four-momenta of the scattered partons from the medium (\textit{recoil}) in the event record, but this can be enabled by a configuration option. We will refer to this mode as `with recoil' and will use this for all the discussions presented in this paper. In addition, \JEWEL{} offers the possibility to output the four momenta of these medium partons before they interacted with the shower, to keep track of total energy and momentum from the medium that is added to the event record \cite{KunnawalkamElayavalli:2017hxo}. This is essential in order to properly subtract the initial thermal momenta of the medium partons. We perform this subtraction using the constituent subtraction procedure~\cite{Berta:2014eza} outlined in Ref.~\cite{Milhano:2022kzx}, before performing the jet reconstruction. 

The sample of $\gamma$+jet events analyzed in this study was generated at $\sqrt{s} = 5.02$ TeV, requiring a prompt photon with $\ptg > 200 \, {\rm GeV}$ at rapidity $\vert \eta^\gamma \vert < 1.44$. The tagged photon is required to be isolated from hadronic activity by demanding the total energy within a cone of radius $R=0.4$ to be less than $5\, {\rm GeV}$. Jets are reconstructed using the anti-kt algorithm~\cite{Cacciari:2008gp} as implemented in {\textsc{Fastjet}}~\cite{Cacciari:2011ma} with a radius parameter of $r_{\rm sub}=0.2$ and $\pt^{\rm sub} > 20$ GeV to suppress background contributions, which form the input for the trimmed large-$R$ jet with $R=1.2$. 
We select the leading $R=1.2$ trimmed jet and require it to have a $\ptjet > 40 \, {\rm GeV}$ within a rapidity of $\vert \eta^J\vert < 2$. For results presented for direct large-$R$ jets in Figs.~\ref{fig:xjg_largeR} and~\ref{fig:girth_largeR}, we  considered the leading $R=1.2$ jet with $\ptjet > 40 \, {\rm GeV}$ while all the photon cuts remain the same as above. The $\PbPb$ samples were generated for $0-10\%$ centrality class with an initial medium temperature of $T_i = 0.55$ GeV. This choice of medium parameters gives an $R_{AA}$ of 0.45-0.55 for $50 < \pt < 350$ GeV jets~\cite{Apolinario:2024apr}.

\subsection{Jet Observables}
Our goal is to study medium-induced broadening using jet substructure observables that encode angular information about the radiation. To  this end, we focus on the width of a jet that can be quantified by the infrared and collinear safe observable \textit{jet girth} $g$, defined as
\begin{equation}
  g = \frac{1}{\pt^{\rm jet}}\sum_{i\in\mathrm{jet}} \pt^i\,\Delta R_{i,\mathrm{jet}}\,,
  \label{eq:girth}
\end{equation}
where $\Delta R_{i,\mathrm{jet}}$ is the distance between particle $i$ and the jet axis, and the sum runs over the jet constituents. The jet girth measures the transverse spread of the radiation in the jet about the jet axis and is particularly sensitive to modifications of wide-angle radiation~\cite{Yan:2020zrz}.

In addition, we find it instructive to also look at the energy distribution as a function of the distance $r$ to the jet axis,
\begin{equation}
    \psi(r) = \frac{\sum_{r_i < r} p_{T,i}}{\sum_{r_i < R} p_{T,i}}\, \quad \rho(r) = \frac{\df \psi(r)}{\df r} \,
\,,\end{equation}
where $\psi(r)$ is the integrated jet profile and $\rho(r)$ is the differential one. The jet girth can be thought of as the first moment of the jet radial profile $\rho(r)$. As a result, $\rho(r)$ provides differential access to angular regions where jet broadening (if any) appears. For the analysis presented here, we use the Winner-Take-All jet axis~\cite{Larkoski:2014uqa} as it is less sensitive to soft radiation effects than the standard jet axis.

\begin{figure}[!t]
    \centering 

    \begin{subfigure}{0.47\textwidth}
        \centering
        \includegraphics[width=\textwidth]{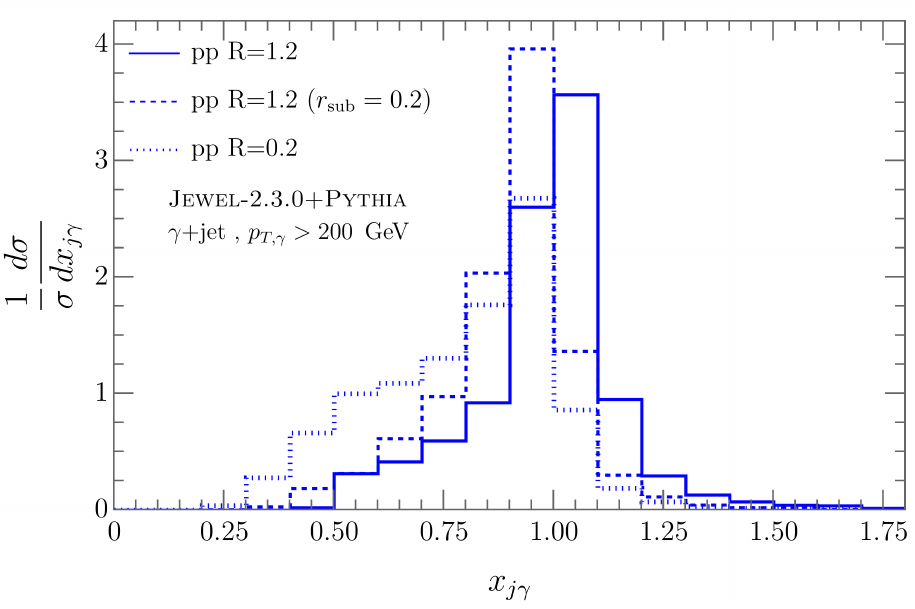}
        \caption{$\pp$ jets}
        \label{fig:xjg_pp}
    \end{subfigure}
    \hfill
    \begin{subfigure}{0.47\textwidth}
        \centering
        \includegraphics[width=\textwidth]{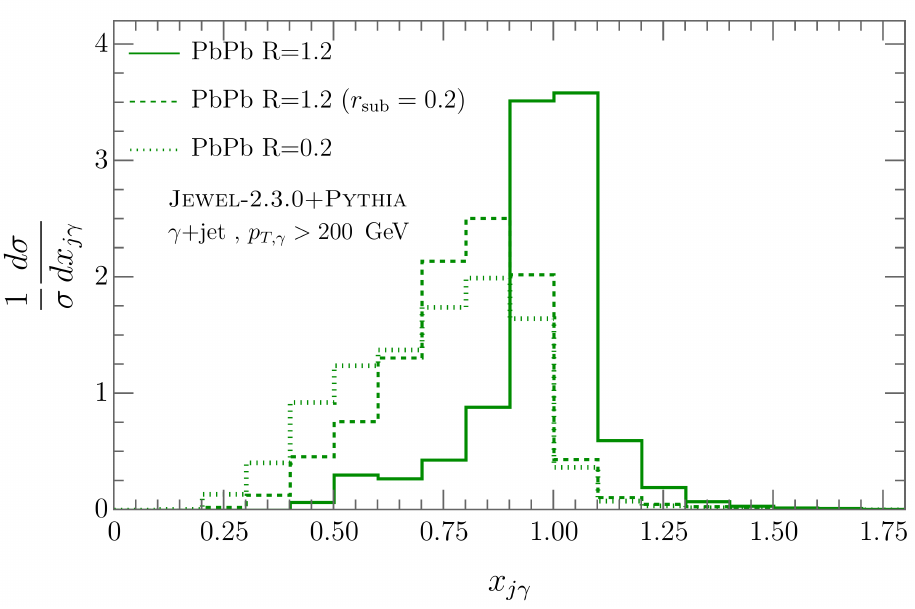}
        \caption{$\PbPb$ jets}
        \label{fig:xjg_PbPb}
    \end{subfigure}

    \caption{The $\xjg$ distribution for direct large-$R$ (solid) jets compared to the $\xjg^{\rm trim}$ of the trimmed jets (dashed) with $r_{\rm sub} = 0.2$ and $p_T^{\rm sub} > 20$ GeV,
    as well as the direct $R=0.2$ (dotted) jets, for (a) $\pp$ and (b) $\PbPb$.
    }    
    \label{fig:xjg_recluster}
\end{figure}

\section{Re-clustering small-$R$ jets as a large-$R$ jet proxy}
\label{sec:reclustering}

As large-$R$ $\gamma$-tagged jets provide the least biased baseline for studying jet substructure modifications, our central goal is to investigate the degree to which the physics encoded in large-$R$ jets can be recovered in an experimentally viable way, by re-clustering a collection of small-radius jets, which we refer to as the trimmed jet. This procedure is similar to the one deployed by ATLAS for inclusive jets in the heavy-ion environment~\cite{ATLAS:2023hso}. 

\subsection{Energy recovered by the trimmed jet}

We first explore to what extent the underlying $x_{j,\gamma}$ distribution can be restored through the trimmed jet, with corresponding $x_{j,\gamma}^{\rm trim}$. Figures~\ref{fig:xjg_pp} and~\ref{fig:xjg_PbPb} show these distributions (dashed curves) for $\pp$ and $\PbPb$ compared to $R=0.2$ (dotted) as well as the $R=1.2$ (solid) radius jets.

\begin{figure}[t]
     \centering
     \includegraphics[scale=0.48]{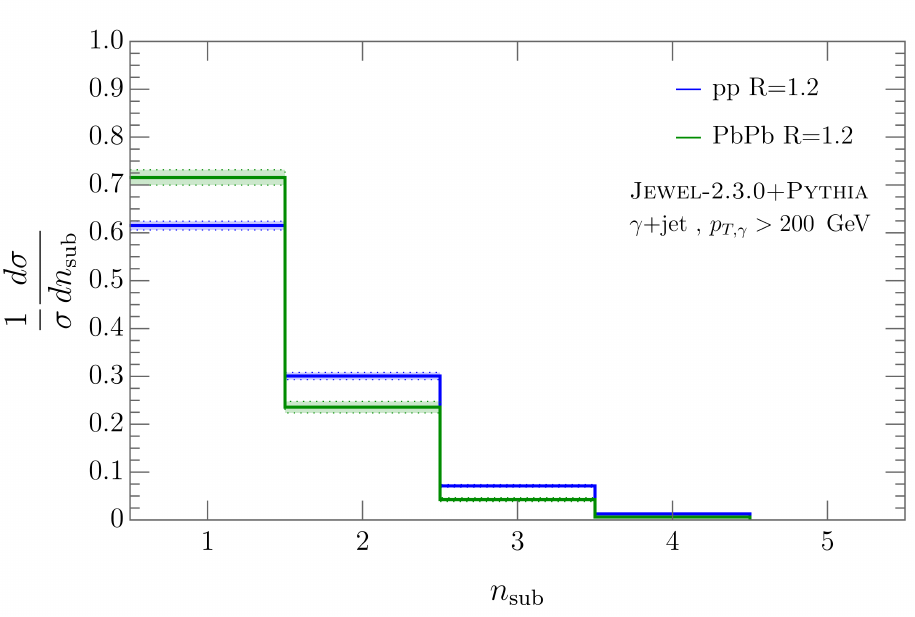}
     \caption{Number of $r_{\rm sub}=0.2$ subjets within the large-$R$ trimmed jet. The fraction of jets with a single subjet is larger in $\PbPb$ than in $\pp$.}
     \label{fig:nsub}
\end{figure}

From the Figure, we observe that the trimming works reasonably well in $\pp$ and brings the $\xjg$ distribution of the trimmed jet closer to 1, i.e., more balanced with the photon than small $R=0.2$ jets.  
Trimmed jets work less well for $\PbPb$ collisions, but still provide a better approximation of large-$R$ jets than an $R = 0.2$ jet, with a somewhat reduced sensitivity of $\xjg$ to energy loss physics.\footnote{As we employ $\pTsub >20 $ GeV, the trimming approach is only useful if the photon transverse momentum is large i.e.~$\ptg >200$ GeV. We have checked that for $\ptg > 100$ GeV, the improvement with $r_{\rm sub}=0.2$ trimmed jets is only marginal. However, this can be compensated by using $r_{\rm sub}=0.4$.} The 
trimmed $\xjg$ distribution in $\PbPb$ is different from $\pp$ due to radiation that falls outside the $r_{\rm sub} =0.2$ core and is below the $\pTsub > 20$~GeV threshold, which is not recovered by the trimming procedure.

To further illustrate the difference between $\pp$ and $\PbPb$, Fig.~\ref{fig:nsub} shows the distribution of the number of $r_{\rm sub} = 0.2$ subjets that are re-clustered within the $R = 1.2$ jet for both $\pp$ and $\PbPb$ jet samples. We observe that in $\PbPb$ collisions, the fraction of single-subjet events is enhanced relative to pp, while the fraction of multi-subjet events is correspondingly suppressed. This observation is consistent with the expectation that subjet structures involving more color charge configurations experience larger quenching than single subjet structures. As a result, the sub-leading subjets are more likely to fall below the $\pTsub > 20$~GeV threshold and be removed from the multi-subjet  sample in $\PbPb$, leading to a suppression of multi-jet structures compared to vacuum. The suppression of multi-subjet configurations in $\PbPb$ naturally limits the amount of energy that can be recovered by the trimming procedure, leading to a larger imbalance with respect to the photon than in $\pp$.

Finally, if one in addition applies grooming~\cite{Larkoski:2014wba} on the subjets of the trimmed jet, the corresponding $\xjg$ distribution is less sensitive to the specifics of the trimming. However, this also weakens the observed broadening which is why we do not investigate it further.\footnote{We still observe a small signal of broadening for the sub-leading jet in 2-subjet configurations when considering the soft drop groomed jet radius $R_g$ as the observable. The mean $R_g$ increases from $0.1155 \pm 0.0005$ in $\pp$ to $0.1191 \pm 0.0013$ in $\PbPb$.}

\subsection{Girth of the trimmed jet}

\begin{figure}[!tb]
    \centering
    \includegraphics[scale=0.48]{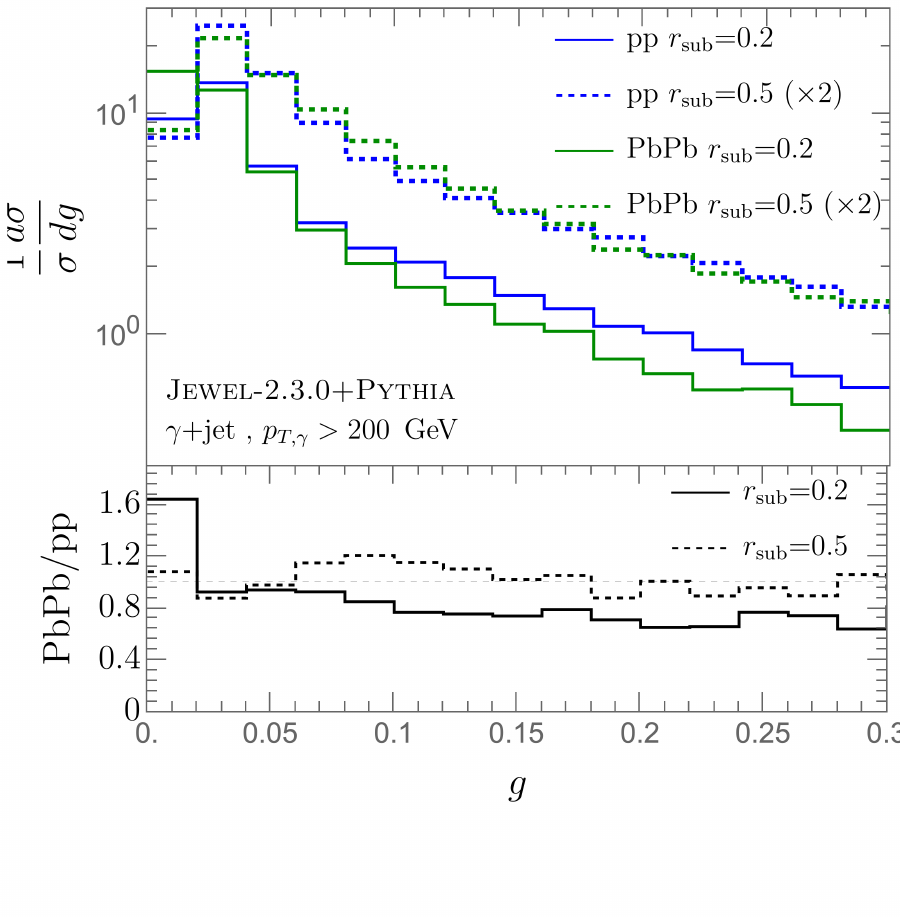}
    \caption{Jet girth computed on the trimmed jet for two choices of subjet clustering radius, $r_{\rm sub} =0.2$ (solid) and $r_{\rm sub} = 0.5$ (dashed), along with the corresponding ratios of $\PbPb/\pp$ in the bottom panel. While no tail is observed at $r_{\rm sub}=0.2$, increasing the subjet clustering radius increases the distribution at large $g$ values relative to $\pp$.} 
    \label{fig:girth}
\end{figure}

As trimmed jets provide a reasonable proxy of large-$R$ jets, one can then use the constituents of the trimmed jet to study the differential pattern of radiation and its re-distribution in the presence of a medium, with the bias due to energy loss reduced. To this end, we analyze the \textit{jet girth} ($g$) of the trimmed jet in Fig.~\ref{fig:girth}, using an inclusive selection on $\xjg > 0.4$.

The girth computed on the trimmed $R=1.2$ jet with $r_{\rm sub} = 0.2$ shows a narrowing of the jet core in $\PbPb$ qualitatively similar to that observed in inclusive jet measurements~\cite{CMS:2013lhm,ALICE:2017nij,CMS:2018zze,ALICE:2018dxf,ALICE:2019qyj,ALICE:2024jtb}, whereas the large-$R$ jet shows a clear broadening, see Fig.~\ref{fig:girth_largeR}.\footnote{Qualitatively similar features are also observed for $\xjg > 0.8$, though it shows a more pronounced narrowing due to the selection bias toward less quenched jets, as discussed in~\ref{sec:selection}.} 
This demonstrates that the medium-induced broadening observed for the direct large-$R$ jet originates largely from radiation that is removed by the trimming procedure, because it lies outside the central $r_{\rm sub}=0.2$ subjet and falls below the $\pt^{\rm sub}>20$ GeV threshold.
Increasing the subjet radius used for trimming to $r_{\rm sub}=0.5$ with the same $\pt^{\rm sub} > 20\, {\rm GeV}$ requirement reveals signals of medium-induced broadening. This shows that medium-induced broadening is mostly visible at relatively low transverse momentum and that relatively large jet cone angles are needed to capture the resulting energy flow at larger angles. 

\begin{figure}[!t]
    \centering 

    \begin{subfigure}{0.47\textwidth}
        \centering
        \includegraphics[width=\textwidth]{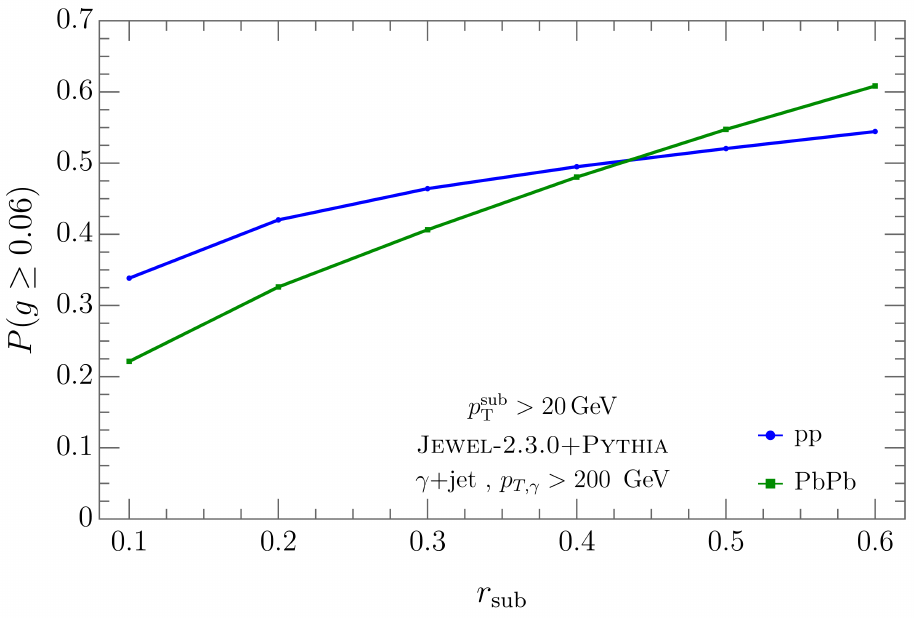}
        \caption{}
        \label{fig:g_R}
    \end{subfigure}
    \hfill
    \begin{subfigure}{0.47\textwidth}
        \centering
        \includegraphics[width=\textwidth]{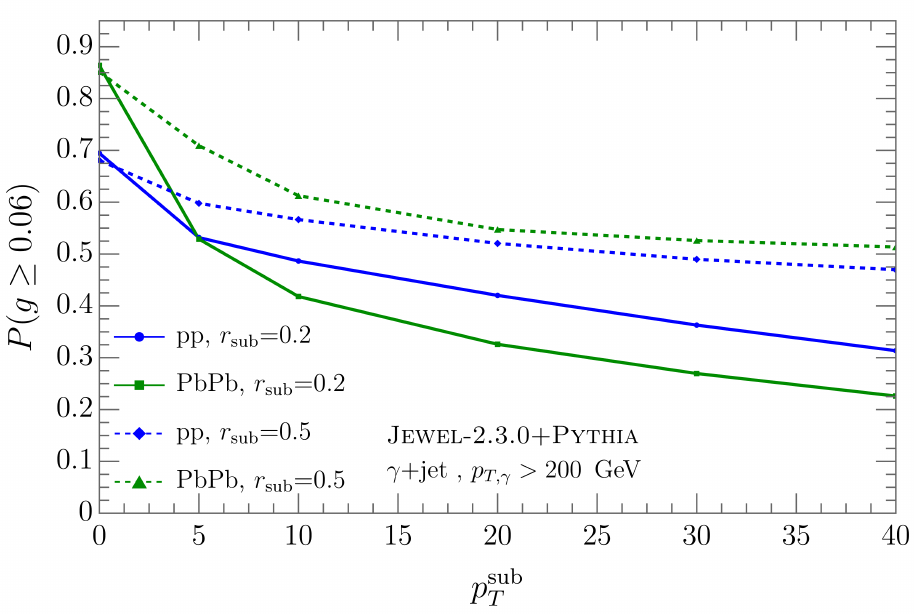}
        \caption{}
        \label{fig:g_pt}
    \end{subfigure}

    \caption{The fraction in the tail $g \geq 0.06$ of the jet girth distribution  as a function of the subjet parameters $r_{\rm sub}$ in (a) and $\pt^{\rm sub}$ in (b).
    }
    \label{fig:girth_tail}
\end{figure}

\begin{figure*}[!t]
    \centering
    \includegraphics[scale=0.45]{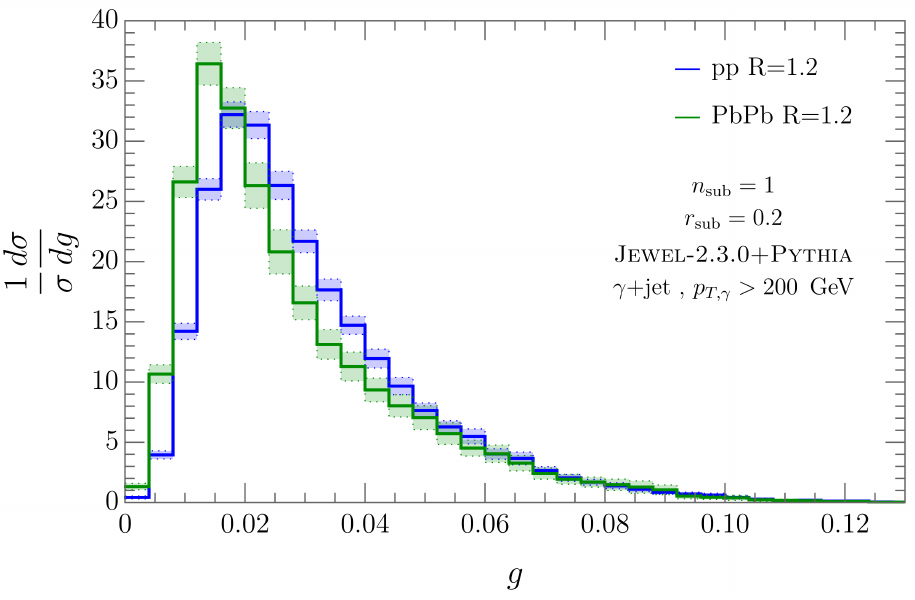}
    \includegraphics[scale=0.45]{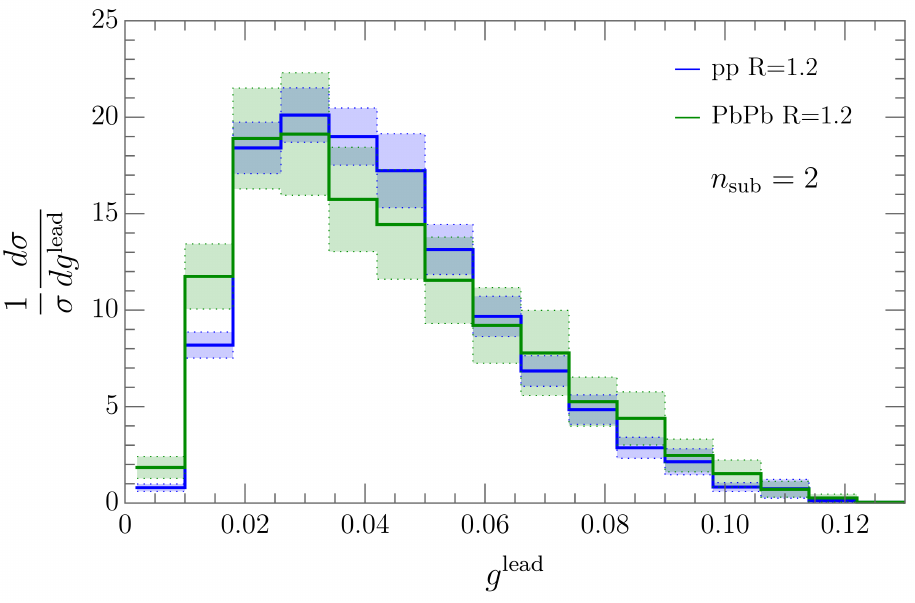}
    \includegraphics[scale=0.45]{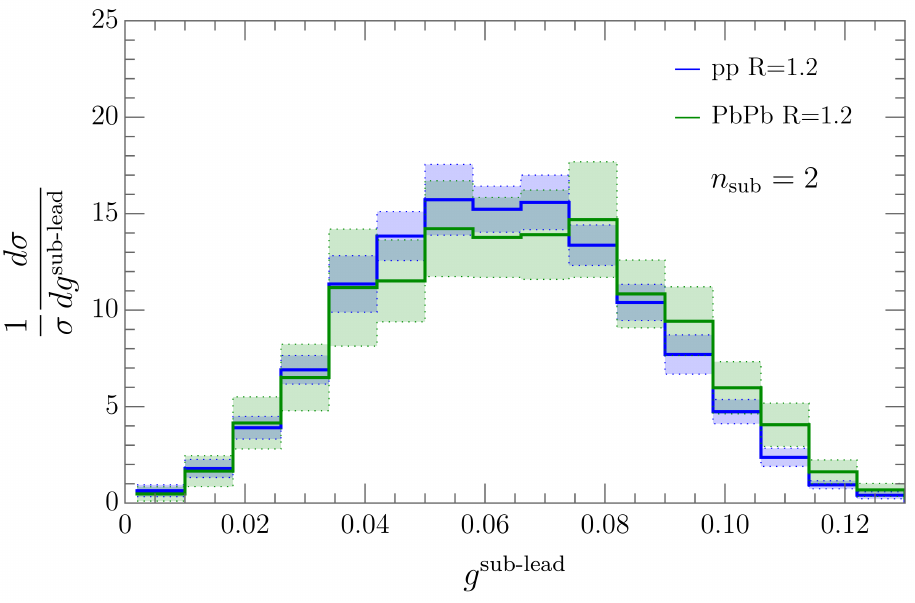}
    \caption{Girth distributions for subjets recoiling against a photon with $p_{T\gamma} > 200$ GeV in \pp{} (blue lines) and PbPb (green curves) events. The three panels show the leading subjet in trimmed jets with only one subjet (left panel) and with two subjets (right panel), as well as the sub-leading subjet (bottom panel). The mean girth values are $0.0294 \pm 0.0002$ ($\pp$) and $0.0263 \pm 0.0004$ ($\PbPb$) for the 1-subjet configuration and $0.0395 \pm 0.0005$ in $\pp$ and $0.0395 \pm 0.0009$ in $\PbPb$ for the leading subjet and $0.0588 \pm 0.0005$ in $\pp$ and $0.0617 \pm 0.0010$ in $\PbPb$ for the sub-leading subjet in the 2-subjet configuration.}
    \label{fig:rz}
\end{figure*}

To systematically explore the modification of the jet girth distribution in the medium, we study its dependence on the subjet radius $r_{\rm sub}$ and minimum threshold $\pt^{\rm sub}$. As the mean and variance of the girth distributions do not sufficiently capture the enhancement of the $\PbPb$ jet distribution in the large $g$ region, we characterize this through the \textit{tail fraction}, \begin{equation}
    P(g \geq g_{\rm cut}) = \sum_{g_i \geq g_{\rm cut}} g_i  \, .
\end{equation}
Here $g_i$ are the normalized counts in the histogram bins of jet girth distribution. 

The resulting tail fraction is shown in Fig.~\ref{fig:g_R} as a function of the subjet radius $r_{\rm sub}$ and as a function of the minimum $\pt$ threshold in Fig.~\ref{fig:g_pt}. From these Figures, we find that the minimum $\pt$ threshold of the subjets seems to have a smaller impact on the trimmed jet's girth distribution compared to the jet radius parameter $r_{\rm sub}$. In particular, for $r_{\rm sub} \gtrsim 0.4$, the enhancement at large $g$ of $\PbPb$ jets becomes visible, indicating that recovering medium-induced broadening requires capturing radiation over sufficiently large angular scales.

\section{Differential analysis of the subjets}
\label{sec:subjets}

We now study the structure of our trimmed jets by separately considering cases where there is only a single $r_{\rm sub} = 0.2$ subjet inside the large-$R$ jet and those containing multiple resolved subjets. This classification provides a natural way to investigate how medium-induced modifications depend on the resolved subjet structure of the jet~\cite{Mehtar-Tani:2024smp,Mehtar-Tani:2025xxd}. In particular, arguments based on color (de)coherence predict that partons separated by an angle larger than the medium resolution angle $\theta_c \sim 1/\sqrt{\hat{q} L^3}$ (where $\hat{q}$ is the jet transport parameter and $L$ is the path length through the medium) lose energy independently~\cite{Mehtar-Tani:2011hma,Casalderrey-Solana:2011ule,Mehtar-Tani:2011vlz,Mehtar-Tani:2017ypq}, while partons separated by an angle smaller than $\theta_c$ lose energy as a single color charge, thereby experiencing overall smaller energy loss. Indications of these differences in energy loss were found by ATLAS~\cite{ATLAS:2023hso,ATLAS:2022vii}, which found a sizable difference in $R_{\rm AA}$ between single and multi-subjet configurations, as well as a dependence on the distance between subjets.

\begin{figure}[!tb]
    \centering
    \includegraphics[scale=0.48]{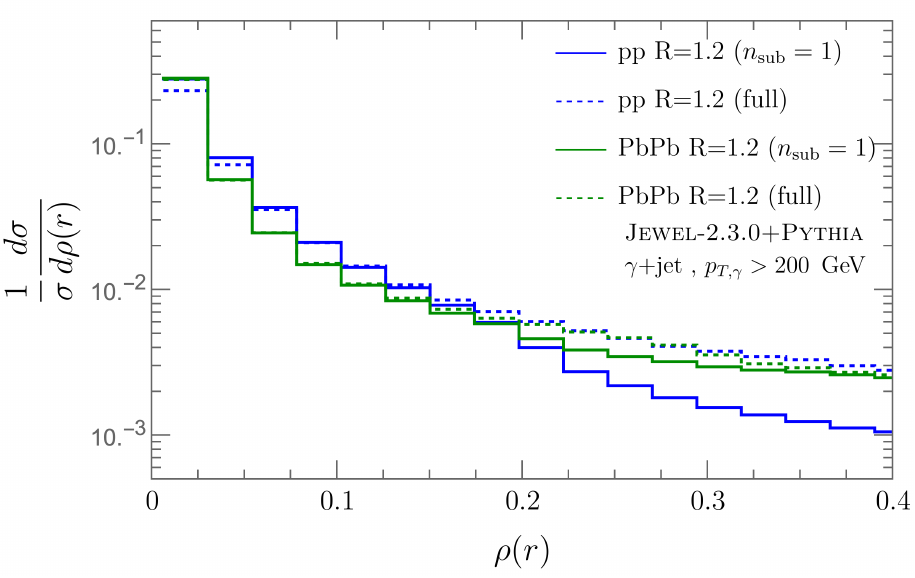}
    \caption{Jet radial profile for $R=1.2$ jets (dashed) and trimmed jets with only a single subjet with radius $r_{\rm sub}=0.2$ (solid). A stronger selection effect is seen for trimmed jets in $\pp{}$ collisions than in \PbPb{} collisions.}
    \label{fig:rho}
\end{figure}

In Fig.~\ref{fig:rz} we study the girth of subjets for the 1-subjet and 2-subjet cases separately. For the 1-subjet configuration, a clear narrowing in $\PbPb$ relative to $\pp$ is observed. While this narrowing looks similar to the jet selection bias, there is no apparent bias in these events: The $x_{j\gamma}$ cut is essentially inclusive in both $\pp$ and $\PbPb$ collisions, see Fig.~\ref{fig:xjg_recluster}, and furthermore the fraction of 1-subjet configuration increases in $\PbPb$ compared to $\pp$, see Fig.~\ref{fig:nsub}. We investigate this further in Fig.~\ref{fig:rho} using the jet radial profile. 

A different picture emerges for the sub-leading subjet in 2-subjet configurations (bottom panel of Fig.~\ref{fig:rz}): while there is a selection effect for sub-leading subjets, i.e.~there are fewer jets with two subjets in PbPb events compared to pp events, the remaining subjets still show a modest broadening. The mean girth increases from $0.0588\pm 0.0005$ in $\pp$ to $0.0617\pm 0.0010$ in $\PbPb$.
We have also verified that the $\xjg$ of the sub-leading subjet is not strongly correlated with the $\xjg$ of the trimmed jets with two subjets, implying that the sub-leading subjet adds non-trivial information. 
 
Finally, the leading subjet in the 2-subjet case (right panel, top row of Fig.~\ref{fig:rz}) shows an intermediate behavior: the mean girth is comparable between pp and $\PbPb$ ($0.0395 \pm 0.0005$ vs.\ $0.0395 \pm 0.0009$), with no significant modification of the shape of the distribution. We also note that the mean value of the girth is larger for the leading subjet in the presence of a second subjet, and even larger for the subleading subjet, which is at least partially due to their lower energies.

To understand the origin of the observed narrowing, we examine the jet radial profile $\rho(r)$ which characterizes the angular distribution of energy in the jet. Figure~\ref{fig:rho} shows the jet radial profile distribution for $\pp$ and $\PbPb$, for large $R=1.2$ jets and trimmed jets with only one subjet of $r_{\rm sub} =0.2$. The $\PbPb$ distribution lies below $\pp$ at small $r$, corresponding to the narrowing observed in the girth of the subjet in Fig.~\ref{fig:rz}, and crosses around $r= r_{\rm sub}$. For larger $r$ the $\pp$ distribution is substantially suppressed, while the $\PbPb$ distribution is much less affected. The single subjet selection has a smaller effect in $\PbPb$ collisions, because the medium-induced radiation is not sufficiently concentrated to be clustered into an energetic subjet. Thus the single subjet selection, combined with a measurement beyond $r_{\rm sub}$, provides a clear way to study the large angle soft radiation characteristic of the medium in $\PbPb$.

\section{Summary \& Outlook}
\label{sec:summary}

We investigated medium-induced jet broadening using $\gamma$+jet events that provide a reduced bias of the jet transverse momentum distribution. 
Using the \JEWEL{} model to generate samples of $\pp$ and $\PbPb$ distributions, we study the jet girth distribution of the large-$R$ jets as a baseline and find a clear signature of in-medium broadening. However, as large-$R$ jets are challenging to measure experimentally, we utilize a re-clustering technique that uses small-$R$ jets with radius $r_{\rm sub}=0.2$ to create a proxy of the $R=1.2$ jet, and refer to it as `jet trimming'.

We first analyzed to what extent the energy loss inherent in small-radius jets is restored through the trimming, comparing it against the benchmark $R=1.2$ jets as well as $R=0.2$ jets to assess the performance of our approach. We find that the trimming approach performs reasonably well with $r_{\rm sub}=0.2$ and $\pt^{\rm sub} > 20$\, GeV for $\ptg >200\, {\rm GeV}$. Given the current experimental reach, lower cuts on $\ptg$ can be considered but require a correspondingly larger $r_{\rm sub}$ and/or smaller $\pt^{\rm sub}$.

We further performed a differential analysis of our trimmed jets to study the internal structure of the 1-subjet and 2-subjet configurations.
Due to energy loss effects, some 2-subjet configurations in $\pp$ end up as single subjets in $\PbPb$ when the $\pt^{\rm sub}$ of the sub-leading subjet falls below the cut, enhancing the fraction of single subjet configurations in $\PbPb$. We find that the jet girth distribution in this case shows a strong narrowing of the jets in $\PbPb$. To understand the origin of this narrowing, we compared the radial profiles of $R=1.2$ jets to those of trimmed jets with a single subjet. Interestingly, for $r>r_{\rm sub}$ the $\pp$ distribution is suppressed due to selecting for one subjet, while the $\PbPb$ distribution is almost unaffected by this. This broadening of $\PbPb$ compared to $\pp$ would not be visible in the girth of the subjet, but can be studied by looking \emph{beyond} the subjet radius.
In the 2-subjet case, we find an observable signal of medium-induced broadening in the girth distribution of the sub-leading subjet.

The analysis presented in this work was carried out in a specific implementation of jet-medium interactions in the \JEWEL{} event generator, and it would be interesting to see whether similar conclusions are reached within other jet quenching models. Analyzing experimental data on $\gamma$+jet events using the approach studied here would present an exciting opportunity to further improve our understanding of jet-medium interactions.

\section*{Acknowledgements}

This work was supported in part by the Dutch Research Council (NWO) as part of the project `Microscopy of the Quark Gluon Plasma using high-energy probes' (VI.C.182.054).

\appendix

\section{Impact of $\xjg$ selection and minimum $\pt$ threshold on the trimmed jets}
\label{sec:selection}

\begin{figure}[!tb]
    \centering
    \includegraphics[scale=0.49]{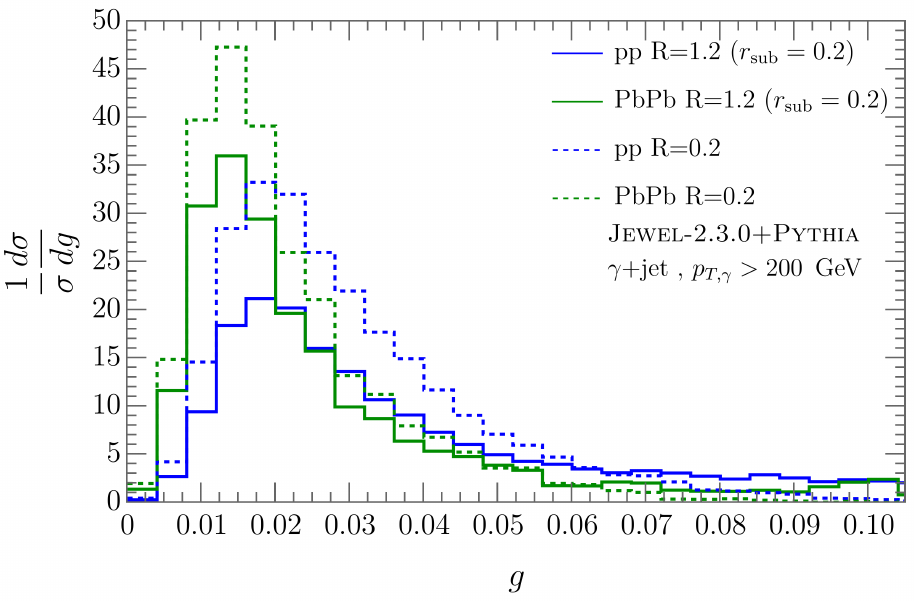}
    \caption{Jet girth distribution of trimmed jets (solid lines) and $R=0.2$ jets (dashed) recoiling a photon with $\ptg > 200$ GeV and $\xjg >0.8$ for $\pp{}$ (blue) and $\PbPb$ (green) collisions.}
    \label{fig:g_xjg8}
\end{figure}

In this appendix, we briefly examine how the girth distribution of the trimmed jet changes as the $\xjg$ selection threshold is varied. Figure~\ref{fig:g_xjg8} shows the girth of trimmed $R = 1.2$ jets for $\xjg > 0.8$ compared to $R=0.2$ jets. The $\xjg >0.8$ selects configurations of jets that are more balanced relative to the photon and as a result have an inherently hard collimated internal structure. Consequently, the narrowing becomes more strongly pronounced compared to the more inclusive jet selection of $\xjg > 0.4$ presented in the main text, see Fig.~\ref{fig:girth}. 
When compared to $R=0.2$ jets (dashed), we find that the trimming procedure reduces this narrowing effect, though the difference is small.

\bibliographystyle{elsarticle-num}
\bibliography{gamjet}
\end{document}